\documentclass[final,5p,times,twocolumn]{elsarticle}

\usepackage{amssymb}

\journal{JHEAP. Swift: 10 years of discovery}

\begin{document}

\begin{frontmatter}



\title{Tidal disruption of stars by supermassive black holes: Status of observations}


\author{S. Komossa}

\address{Max-Planck-Institut f{\"u}r Radioastronomie, Auf dem H{\"u}gel 69, 53121 Bonn, Germany}

\begin{abstract}
Stars in the immediate vicinity of supermassive black holes (SMBHs) can be ripped
apart by the tidal forces of the black hole. The subsequent accretion of the stellar
material causes a spectacular flare of electromagnetic radiation.
Here, we provide a review of the observations of tidal disruption events (TDEs),
with an emphasis on the important contributions of \textit{Swift} to
this field.
TDEs represent a new probe of matter under strong gravity,
and have opened up a new window into studying accretion physics under extreme conditions. The
events 
probe relativistic effects, provide a new means of measuring black hole spin, and represent signposts
of intermediate-mass BHs, binary BHs and recoiling BHs.  
Luminous, high-amplitude X-ray flares, matching key predictions of the tidal disruption
scenario, have first been discovered with \textit{ROSAT},
and more recently with other missions and in other wavebands. 
The \textit{Swift} discovery of two $\gamma$-ray emitting, jetted TDEs,
never seen before, has provided us with a unique probe of the early
phases of jet formation and evolution, and Swift\,J1644+75 has
the best covered lightcurve of any TDE to date.
Further, \textit{Swift} has made important contributions in providing
well-covered lightcurves of TDEs discovered with other instruments, setting
constraints on the physics that govern the TDE evolution, and including the discovery
of the first candidate binary SMBH identified from a TDE lightcurve.

\end{abstract}


\end{frontmatter}


\section{Introduction}
\label{}

The tidal disruption of stars is an inevitable consequence of the
presence of supermassive black holes (SMBHs) at the cores of galaxies.  
Stars in close approach  
can be ripped apart by the tidal forces of the SMBH.
A significant fraction of the stellar material will subsequently be accreted,
producing a luminous flare of electromagnetic radiation.

The occurrence of TDEs was first predicted based on theoretical 
considerations in the seventies. 
At that time, many important theoretical foundations were laid,  
and first ideas emerged,
on the relevance and usage of TDEs in astrophysical context (Hills 1975, Frank \& Rees 1976, 
Young et al. 1997, Kato \& Hoshi 1978, Gurzadian \& Ozernoi 1979, 
Carter \& Luminet 1982, Luminet \& Marck 1985, Rees 1988).
Questions which were raised included: 
Do TDEs
provide a fuel source for quasars?
How does the occurrence of TDEs  
constrain the presence of a SMBH at our Galactic center? 
Can extreme TDEs explain variants of gamma-ray bursts (GRBs)?  
Are TDEs a  unique signpost of the presence of dormant SMBHs at the cores of
inactive galaxies ? 

Events were discovered in the nineties in form of luminous, soft X-ray outbursts
from otherwise quiescent galaxies during the \textit{ROSAT} all-sky survey, and
have more recently been found with other X-ray missions, and at longer wavelengths. 
The \textit{Swift} mission has been a 
game changer in this field. 
Its discovery of the first TDE which launched a relativistic jet, Swift\,J1644+57, has triggered
many theoretical studies on the formation of radio jets,
and Swift\,J1644+57 has now the best-covered lightcurve of any TDE to date. Further, \textit{Swift} has
been used for follow-ups of TDEs discovered by other missions, providing excellent lightcurves
and contributing to the discovery of the first candidate binary SMBH in a quiescent galaxy.

A star
is  disrupted, once the tidal forces of the SMBH 
exceed the self-gravity of the star
(Hills 1975).
The distance at which this happens, the tidal radius, is given by
\begin{equation}
r_{\rm t} \simeq 7\,\times\,10^{12}\,{\bigg{(}}{M_{\rm BH}\over {10^{6} M_\odot}}{\bigg{)}}^{1 \over 3}
    {\bigg{(}}{M_{\rm *}\over M_\odot}{\bigg{)}}^{-{1 \over 3}} {r_* \over r_\odot}~{\rm cm}\,. 
\end{equation}
A fraction of the stellar material
will be on unbound orbits and escape, while the rest will
eventually be accreted (Fig. 1).  The events appear as luminous transients. 
Their emission peaks in the UV or soft X-rays, declining on the timescale of months to years
(e.g., Rees 1990, Evans \& Kochanek 1989, Cannizzo et al. 1990). 
Because tidal radius and Schwarzschild radius depend differently on BH mass,
solar type stars are swallowed whole for SMBH masses exceeding $\sim$ 10$^{8}$ M$_{\rm \odot}$
(SMBH spin can raise this limit; Beloborodov et al. 1992). White dwarfs can be tidally
disrupted for SMBH masses below $\sim$ 10$^{5}$ M$_{\rm \odot}$.   

\begin{figure*}[t]
\begin{center}
 \includegraphics[width=12cm]{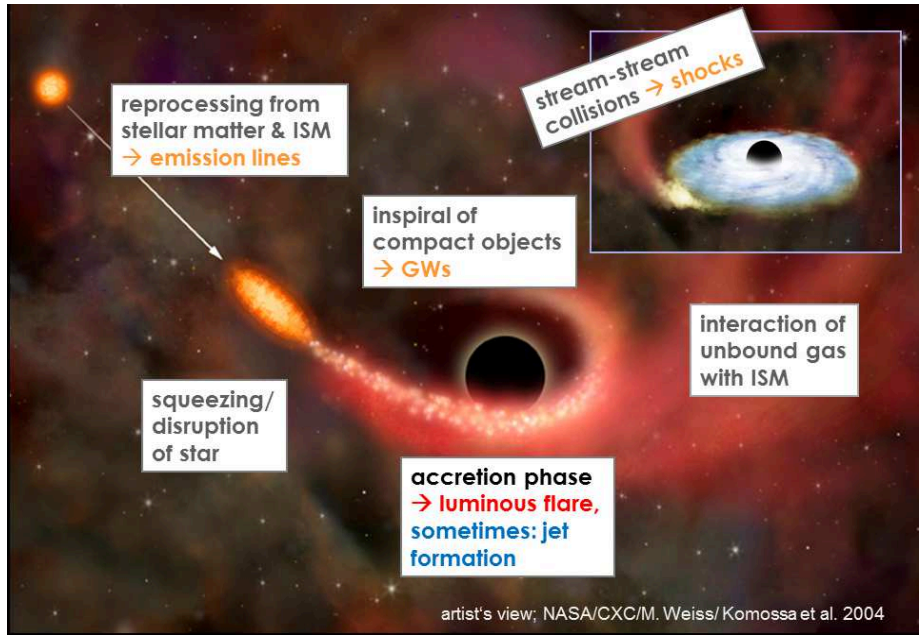}
\caption{Sites and sources of radiation during the evolution of a TDE. In
most cases, the accretion phase is the most luminous electromagnetic phase.
Only some TDEs launch radio jets. 
}
\end{center}
\end{figure*}

Computations of the stellar evolution, including the early stages of stellar deformation,
the actual disruption, accretion and ejection of material, the development
of a disk wind, and the evolution and 
late stages of the accretion phase, and for different types of stars,
are very challenging. 
Recent state of the art modelling
has addressed the different stages of TDE evolution under various conditions
(for the most recent results, see, e.g., Bonnerot et al. 2015, Guillochon \& Ramirez-Ruiz 2015,
Hayasaki et al. 2015, Miller 2015, Piran et al. 2015, Shiokawa et al. 2015,  
and references therein). 

Other recent studies have addressed the disruption rates for different galaxy morphologies
(e.g., Brockamp et al. 2011, Vasiliev 2014, Stone \& Metzger 2014, Zhong et al. 2014),
and for recoiling and binary  SMBHs
(e.g., Komossa \& Merritt 2008, Chen et al. 2009, 2011, Stone \& Loeb 2011, 2012a, 
Li et al. 2012, Liu \& Chen 2013)  
and spinning BHs (Kesden 2012).  
These have shown that rates are strongly boosted in some phases of binary SMBH
evolution and the early phase of SMBH recoil, and that rates strongly depend on BH spin
for the most massive BHs ($M>10^8$ M$_{\rm \odot}$).   
Tidally disrupted stars will also produce a gravitational wave signal along with the electromagnetic
emission (e.g., East 2014, and references therein). 

There is now increasing evidence that SMBHs reside at the centers of many
massive galaxies (review by Graham 2015). TDEs likely significantly contributed to growing SMBHs
at low masses ($M_{\rm BH} < 10^{5-6}$ M$_{\odot}$; Freitag \& Benz 2002), 
while stars swallowed
whole may have contributed to SMBH growth at high masses
(Zhao et al. 2002).   
One initial idea was to use TDE flares as tracers of dormant SMBHs in 
quiescent galaxies (e.g., Rees 1988). This will continue 
to be an important topic in the future,
since the luminous accretion flares reach out to large cosmic distances,
and TDEs will be detected in large numbers in future transient surveys (Sect. 8).  
In addition, many other applications have been suggested, 
making use of the characteristic TDE properties and rates: 

\begin{itemize} 

\item In X-rays, TDEs probe relativistic effects (via emission-line profiles or 
precession effects in the Kerr metric) and  
the extremes of accretion physics at high rates and near the last stable orbit,
and provide us with a new means of measuring BH spin.  

\item Jetted TDEs provide new insight into the formation and early evolution of radio jets, 
and may shed new light on related issues like the cause of the radio-loud radio-quiet dichotomy
of active galactic nuclei (AGN).   

\item TDEs, once detected in large numbers, will unveil the population of IMBHs in the universe. 

\item TDE rates depend on, and therefore trace, stellar dynamics in galaxy cores 
on spatial scales which cannot 
be resolved directly.  

\item TDEs are signposts of binary SMBHs and recoiling BHs, 
because their rates are strongly enhanced
under these conditions, and TDEs will occur off-nuclear if the SMBH is recoiling.  

\item TDEs in gas-rich environments will illuminate the circum-nuclear material, so that the reprocessed
   emission lines and their temporal evolution provide us with 
   an unparalleled opportunity of reverberation mapping 
   the cores of quiescent galaxies.  

\end{itemize} 

Here, we present an overview of the status of observations of TDEs, 
highlighting the important role plaid by the \textit{Swift} mission
(Gehrels et al. 2004, Gehrels \& Cannizzo 2015).
An accompanying review by G. Lodato (2015) will focus on theoretical aspects of tidal
disruption.  

\begin{figure}[t]
 \includegraphics[width=9.0cm]{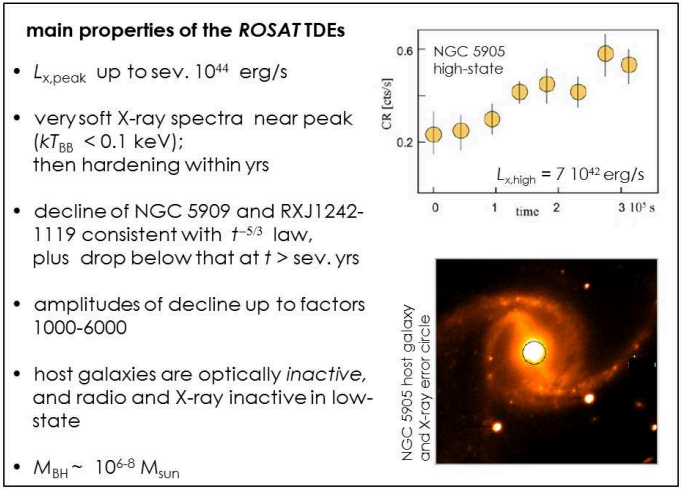}
\caption{Summary of the main properties of the \textit{ROSAT} events. The figures show
the rise to the highest observed state of NGC\,5905 during the RASS and an image of the host galaxy.} 
\end{figure}

\begin{figure}[ht]
 \includegraphics[width=9.5cm]{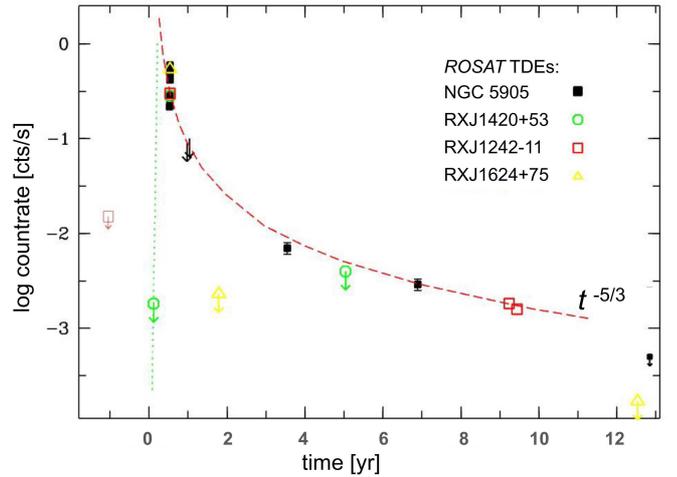}
\caption{Joint X-ray lightcurve of the \textit{ROSAT} TDEs, all shifted to the same
peak time.
The decline is consistent with a $t^{-5/3}$ law (dashed lined).
This point was first made based on \textit{ROSAT} data of NGC\,5905
(Komossa \& Bade 1999), and later for the overall luminosity evolution
of the sources displayed above (e.g., Fig. 1 of Komossa 2004).
RXJ1242-1119 shows a further drop in X-rays at late times (not shown here), deviating from the early
phase decline law, implying a total amplitude of decline of a factor $\sim$1000 (Komossa 2005).
}
\end{figure}

\section{TDEs in soft X-rays (non-jetted)}

\subsection{\textit{ROSAT} TDEs}

The {\it{ROSAT}} observatory (Tr\"umper 2001) with its high sensitivity, long lifetime,
all-sky
coverage in its first year of operation, and soft X-ray response (0.1--2.4 keV), 
was well suited for the detection of TDEs.  
Luminous, high-amplitude
X-ray flares from quiescent galaxies, matching all order-of-magnitude predictions of the tidal 
disruption scenario (e.g., Rees 1990), have first been
discovered during the \textit{ROSAT} X-ray all-sky survey (RASS; Fig. 2, 3).  
Four main events were identified, from the galaxies NGC\,5905 (Bade et al. 1996,
Komossa \& Bade 1999; see also Li et al. 2002), RXJ1242$-$1119 (Komossa \& Greiner 1999),  
RXJ1624+7554 (Grupe et al. 1999), and RXJ1420+5334 (Greiner et al. 2000). 
Among these, NGC\,5905 and RXJ1242$-$1119  are the best-covered events in terms of their {\em long-term}
X-ray lightcurves, spanning time intervals of more than a decade, with amplitudes of 
decline {\em{larger than a factor of 1000}} 
 (Komossa et al. 2004,
Halpern et al. 2004, Komossa 2005).

NGC\,5905 was first noticed due to its luminous, soft ($kT=0.06$ keV) X-ray emission 
with peak luminosity in the soft X-ray band of $L_{\rm x,peak} = 7\,10^{42}$ erg/s
during the RASS. It remained bright
for at least $\sim$5 days (the time interval its position was repeatedly
scanned during the RASS) increasing in
luminosity to the observed peak. X-rays then declined on the timescale
of months to years (Fig. 3). Within the errors, the X-rays came from the center of
this nearby barred spiral galaxy ($z=0.011$; Fig. 2).
While the X-ray spectrum was initially very soft, it had hardened
significantly ($\Gamma_{\rm x} = -2.4$)
3 years later, when re-observed with \textit{ROSAT}.
The decline of its X-ray lightcurve is well consistent with the predicted  
$t^{-5/3}$ law, as first reported
based on its \textit{ROSAT} observations (Komossa \& Bade 1999) and
confirmed with \textit{Chandra} (Halpern et al. 2004).
All observations of this event are
in very good agreement with tidal disruption theory (Bade et al. 1996, Komossa \& Bade 1999).

Whenever enough data exist, the \textit{ROSAT} events,
and most of the more recent soft X-ray TDEs (next Section), follow
a similar trend in spectral and lightcurve evolution as NGC\,5905,
providing independent evidence that  the same mechanism was at work
in all cases.


\subsection{New soft X-ray TDEs and \textit{Swift} follow-ups}

More recently, similar X-ray events have
been detected with \textit{Chandra} and \textit{XMM-Newton}, based on dedicated searches or
serendipitous discoveries. 
The \textit{XMM-Newton} slew survey has been used to identify new bright TDEs based on a comparison
with the \textit{ROSAT} data base, and a few events have been found so far (Esquej et
al. 2007, 2008, Saxton et al. 2012, Saxton et al. 2015 in prep). 
Among these, SDSSJ120136.02+300305.5 has the best-covered first-year lightcurve
(Saxton et al. 2012), based on follow-ups with \textit{XMM-Newton} and \textit{Swift}.
Overall, the X-rays continue fading after high-state. Additional large-amplitude
variability is apparent on the timescale of weeks (Fig. 4).
The X-ray spectrum of SDSSJ120136.02+300305.5, 
observed with \textit{XMM-Newton} weeks and months after high-state
is very soft (no photons detected beyond 2-3 keV), but is not well fit with
black-body emission. It is consistent with a broken powerlaw or a Bremsstrahlung-like
spectral shape.
 
A few TDEs were identified in clusters of galaxies (Cappelluti et al. 2009, Maksym et al. 2010, 2013,
Donato et al. 2014). The most likely counterpart of the source WINGS\,J1348 
in Abell 1795 is a dwarf galaxy, and the disrupting black hole is of 
relatively low mass, $M_{\rm BH} < 10^6$ M$_{\rm \odot}$ (Maksym et al. 2013, 2014a, Donato et al. 2014). 
A second candidate TDE hosted by a dwarf galaxy was reported by Maksym et al. (2014b).  
  
Other events emerged through systematic searches of the \textit{XMM-Newton} data base (Lin et al. 2011, 2015) 
and new searches of the \textit{ROSAT} data base (Khabibullin \& Sazonov 2014, Maksym et al. 2014b).
The events cover X-ray luminosities in the range (10$^{42}$ -- several 10$^{44}$) erg/s, and arise in
relatively nearby galaxies ($z=0.03-0.2$) which are optically quiescent (i.e., they lack the characteristic
optical narrow emission lines of AGN).  
The \textit{Swift} mission has been essential in providing rapid follow-ups of several of these events, 
confirming the fading X-rays, and providing tight constraints on the luminosity evolution. 

Overall, the salient properties of the soft X-ray TDEs detected with \textit{ROSAT}, 
\textit{XMM-Newton} and \textit{Chandra}
can be summarized as follows: 

\begin{itemize}

\item Peak luminosities are large, up to several 10$^{44}$ erg/s in the soft X-ray band.

\item Amplitudes of decline reach factors up to $1000-6000$ (the \textit{ROSAT} events), 
more than a decade after
 the observed high-states.

\item X-ray spectra are very soft during the
high-states ($kT_{\rm BB} \sim 0.04-0.1$ keV),
followed by a spectral hardening on the time scale of years.

\item Host galaxies show essentially no evidence for {\em permanent} activity
as it is seen in AGN.
Years after the flare (and before, when data exist), 
the galaxies are optically inactive, radio inactive, and X-ray
inactive.

\item X-ray lightcurves 
decline on the timescale of months--years,
and are overall consistent with the law $L \propto t^{-5/3}$ predicted
by the fall-back model of tidal disruption theory.

\item SMBH masses (derived from applying scaling relations or multi-wavelength correlations) 
are mostly on the order of 10$^6 - 10^8$ M$_{\rm \odot}$. 

\item Only a small amount of the stellar material needed to be accreted to power the
observed emission, typically $<$ 10\% M$_{\odot}$.

\end{itemize}

\subsection{A candidate binary SMBH} 

The lightcurve of the TDE from SDSSJ120136.02+300305.5 (Sect. 2.2, Fig. 4) does show the
overall downward trend expected after tidal disruption. However, one month after the peak,
the X-ray emission suddenly dropped by a factor $>50$ within a week and the source was no longer detected
by \textit{Swift}.  X-rays re-appeared after 115d, and then dropped a second time.

If the intermittence of the X-ray emission was due to absorption, the first 
disappearance in X-rays would require a huge
column density of gas of order $N_{\rm H} > 10^{23}$ cm$^{-2}$.
While in AGN, a broad-line cloud could be the source of such
absorption, the optical spectrum of SDSSJ120136.02+300305.5 is that of a non-active galaxy,
without any detectable emission lines neither before nor after the outburst. The chances of molecular
clouds in the very core repeatedly crossing our line-of-sight toward the X-ray source are
therefore very low. Further, no jet was launched by this TDE, since no radio emission was detected at all
(Saxton et al. 2012).

However, the characteristic intermittence and recovery of the lightcurve of SDSSJ120136.02+300305.5 is
reminiscent of predictions by Liu et al. (2009), who computed TDE lightcurves in binary SMBHs.
In that case, the second SMBH acts as a perturber and temporarily interrupts the accretion stream
on the primary. Simulations by Liu et al. (2014) have shown, that the lightcurve of
SDSSJ120136.02+300305.5 is well consistent with a binary SMBH model (Fig. 4) with a primary
mass of $10^6$ M$_{\odot}$, a mass ratio $q \sim 0.1$ and semi-major axis of 0.6 milli-pc (mpc).

This is the first supermassive binary BH (SMBBH) candidate identified in a non-active host galaxy, 
and the one with the most compact orbit
among the known SMBBH candidates (review by Komossa \& Zensus 2015). It has overcome the 
``final parsec problem'' (e.g., Colpi 2014). Upon coalescence,
it will be a strong source of gravitational wave  emission in the sensitivity range of space-based
gravitational wave detectors.  If significant numbers of SMBBHs exist at the cores of non-active
galaxies, we expect to see more such events in \textit{Swift} lightcurves of TDEs. A good lightcurve coverage
is essential for constraining the system parameters.

\begin{figure}[t]
 \includegraphics[width=8.9cm]{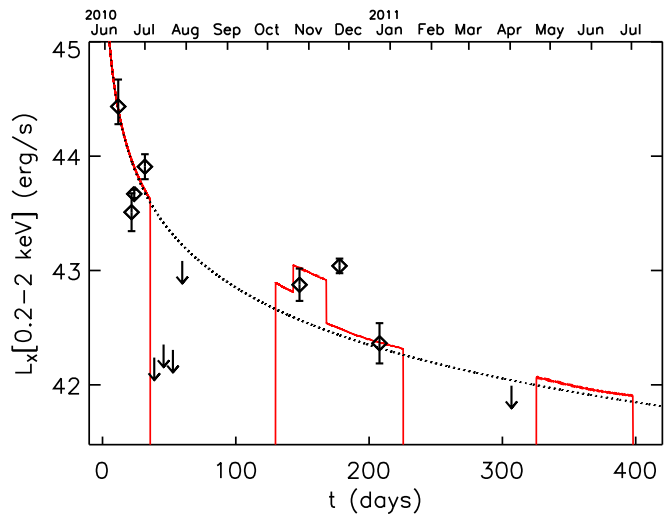}
\caption{\textit{Swift} and \textit{XMM-Newton} X-ray lightcurve of the X-ray 
outburst from SDSSJ120136.02+300305.5,
and predictions from the SMBBH model
of Liu et al. (2014, red solid line).
}
   \label{fig3}
\end{figure}

\section{UV and optical flares}

The \textit{Galaxy Evolution Explorer} (\textit{GALEX}; Martin et al. 2005) mission
performed the first sky survey in the UV from
space, and was well suited for a search of  
TDEs at UV wavelengths.  
Three events were first identified in the UV.  The first
one was detected in the \textit{GALEX} Groth field and showed UV variability 
by a factor of a few (Gezari et al. 2006).
An elliptical galaxy at redshift $z$=0.370 was identified as host.
Its spectrum shows only faint emission from [OIII] and is dominated by stellar
absorption features.   
Two more events were identified with \textit{GALEX} (Gezari et al. 2008, 2009; our Tab. 1).  
 
In the optical bandpass, several photometric sky surveys have become online recently,
performing wide-field transient surveys at high cadence [e.g., PTF (Law et al. 2009),
Pan-STARRS (Kaiser et al. 2002), and for selected areas also SDSS (York et al. 2000)].
These have found several TDE candidates based on their optical
variability, with lightcurves (or spectra) which look different from those of known supernovae.

The first event identified optically was from SDSS, associated
with the galaxy SDSSJ095209.56+214313.3 (Komossa et al. 2008).  
It shows only mild optical variability of $\sim$1 mag,
but was noticed for its transient optical emission lines which imply a luminous high-energy
flare (Sect. 4). Van Velzen et al. (2011) performed a search
for optical flares using the SDSS stripe 82 data base, and presented two probable TDEs, optically variable
by factors 2 -- 2.5. One of them showed evidence for variable broad-line emission.   
Further optical flares, some followed up with \textit{Swift UVOT},  
were reported by Cenko et al. (2012a),
Gezari et al. (2012), 
Holoien et al. (2014), Vinko et al. (2015), Arcavi et al. (2014), and Chornock et al. (2014). Several
of them come with variable optical emission lines (further discussed in Sect. 4),
while the spectrum of PS1-11af (Chornock et al. 2014) is featureless except for two deep 
UV absorption features. 
 
Overall, the UV and optical events have spectral energy distributions 
that are peaked at much lower temperatures
($T_{\rm bb} \approx 10^4$ K) than the X-ray events ($T_{\rm bb} \approx 10^5$ K), and
the majority of the low-temperature events did not have detectable X-ray emission  (even though
for the higher-redshift UV sources limits on X-ray emission, including in low-state,
are not as tight as those for the more 
nearby \textit{ROSAT}, \textit{Chandra} and \textit{XMM-Newton} X-ray events).  

\section{Emission-line transients} 

When travelling through any gaseous material of the galaxy
core, the bright EUV-X-ray continuum of a TDE is 
reprocessed into emission lines. 
These emission lines provide us a with a powerful new opportunity
of measuring the physical conditions in the circum-nuclear material.

When the luminous electromagnetic radiation
travels across the galaxy core, it will photoionize
any circum-nuclear material (Rees 1998, Ulmer 1999)
and  the tidal debris itself 
(Bogdanovic et al. 2004, Strubbe \& Quataert 2011) and is reprocessed into
line radiation. This
emission-line signal enables us to
perform reverberation mapping of any gaseous material
in the galaxy core, including, in principle, 
high-density gas like the broad-line region
(if present), the molecular torus,
and the interstellar medium. We will also  be able  
to address, which of these regions are permanently present in non-active galaxies.
The emission-line fluxes, line widths, line shifts, and their
evolution with time, tightly constrain the amount, density, composition,
dynamics and geometry of the circum-nuclear material.

\begin{figure}[t]
\includegraphics[width=8.9cm]{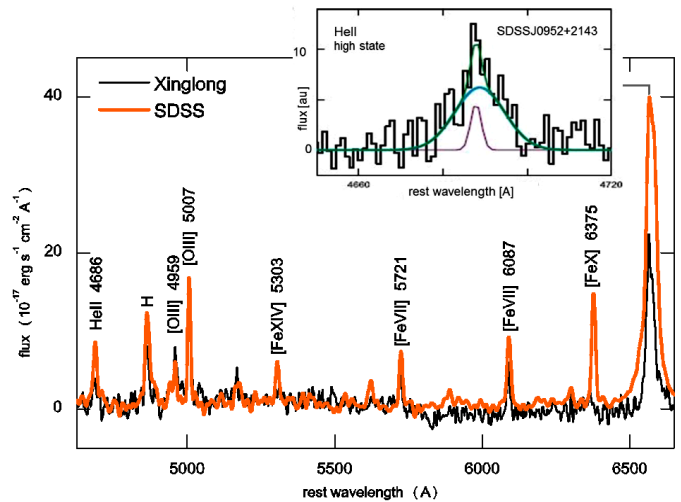}
\caption{Strong and transient Balmer and high-ionization emission lines in the optical
spectrum of SDSSJ\,095209.56+214313.3 (note that the resolution of the SDSS spectrum was
decreased for this plot, to match that of the {\it{Xinglong}} spectrum). The inset
shows a zoom on HeII\,$\lambda$4686.  
While the highest-ionization
iron lines were very bright initially (SDSS spectrum, red),
they were much fainter in a
{\it{Xinglong}} spectrum (black) taken several years later 
(Komossa et al. 2008; see Wang et al. 2012 for
several more cases). For examples which lack the narrow, high-ionization lines,
but still show bright and broad emission of Balmer lines and/or HeII, see,
Wang et al. 2011 (their Fig. 1), Gezari et al. 2012 (their Fig. 1), and  
Arcavi et al. 2014 (their Fig. 4). 
}
\end{figure}

Recently, large optical spectroscopic surveys (SDSS), and follow-ups
of sources identified in photometric transient surveys (PanSTARRS, PTF, ASAS-SN)  
have enabled the discovery of
several well-observed cases of transient
optical emission lines, arising from otherwise
quiescent galaxies (as judged from their low-ionization emission-line ratios;
or from the absence of any narrow emission lines). 
All of them exhibit bright, broad, fading emission from helium and/or hydrogen
(Komossa et al. 2008, 2009,
Wang et al. 2011, 2012, van Velzen et al. 2011, 
Gezari et al. 2012, Gaskell \& Rojas Lobos 2014, Holoien et al. 2014,
Arcavi et al. 2014), while
some of them additionally show transient super-strong iron coronal lines
up to ionization stages of Fe$^{13+}$ including transitions of [FeVII], [FeX], and
[FeXIV]
(Komossa et al. 2008, 2009, Wang et al. 2011, 2012, Yang et al. 2013).
All events have been interpreted as candidate TDEs{\footnote{All of
these events are consistent with arising from the galaxy cores, and look significantly
different from known supernovae (spectra and/or lightcurves). 
Nonetheless, an SN origin cannot be
ruled out yet, and SNe exploding in the dense, gas-rich core region of galaxies
may well look different from other SNe (see the discussion by Komossa et al. 2009,
Drake et al. 2011, and Arcavi et al. 2014). 
The high inferred X-ray luminosities from the strong coronal lines place
particularly tight constraints on any supernova origin
(Komossa et al. 2009).}}.   

\begin{table*}
\begin{center}
\begin{tabular}{lcll}
\hline
source name & redshift & discovery mission & reference \\
\hline
\hline
\multicolumn{4}{c}{soft X-ray events} \\
\hline
NGC\,5905     & 0.011 & \textit{ROSAT} & Bade et al. 1996, Komossa \& Bade 1999 \\
RX\,J1242--1119 & 0.050 & \textit{ROSAT} & Komossa \& Greiner 1999 \\
RX\,J1624+7554  & 0.064 & \textit{ROSAT} & Grupe et al. 1999 \\
RX\,J1420+5334  & 0.147 & \textit{ROSAT} & Greiner et al. 2000 \\
NGC\,3599   & 0.003 & \textit{XMM-Newton} & Esquej et al. 2007, 2008 \\
SDSS\,J1323+4827  & 0.087 & \textit{XMM-Newton} & Esquej et al. 2007, 2008 \\
TDXF\,1347--3254  & 0.037 & \textit{ROSAT}  & Cappelluti et al. 2009 \\
SDSS\,J1311-0123  & 0.195  & \textit{Chandra} & Maksym et al. 2010 \\
2XMMi\,1847--6317 & 0.035 & \textit{XMM-Newton} & Lin et al. 2011\\
SDSS\,J1201+3003  & 0.146 & \textit{XMM-Newton} & Saxton et al. 2012 \\
WINGS\,J1348      & 0.062 & \textit{Chandra} & Maksym et al. 2013, Donato et al. 2014\\
RBS1032           & 0.026 & \textit{ROSAT} & Maksym et al. 2014b, Khabibullin \& Sazonov 2014 \\
3XMM\,J1521+0749  & 0.179 & \textit{XMM-Newton} & Lin et al. 2015 \\
\hline
\multicolumn{4}{c}{hard X-ray events} \\
\hline
Swift\,J1644+57 & 0.353 & \textit{Swift} & Bloom et al. 2011, Burrows et al. 2011, \\
                &       &       & Levan et al. 2011, Zauderer et al. 2011\\
Swift\,J2058+0516 & 1.186 & \textit{Swift} & Cenko et al. 2012b\\
\hline
\multicolumn{4}{c}{UV events} \\
\hline
J1419+5252 & 0.370  & \textit{GALEX} & Gezari et al. 2006 \\ 
J0225--0432& 0.326  & \textit{GALEX} & Gezari et al. 2008 \\
J2331+0017 & 0.186  & \textit{GALEX} & Gezari et al. 2009  \\ 
\hline
\multicolumn{4}{c}{optical events} \\
\hline
SDSS\,J0952+2143$^1$ & 0.079 & SDSS & Komossa et al. 2008\\
SDSS\,J0748+4712$^1$ & 0.062 & SDSS & Wang et al. 2011 \\
SDSS\,J2342+0106   & 0.136 & SDSS & van Velzen et al. 2011 \\
SDSS\,J2323--0108  & 0.251 & SDSS &  \\
PTF\,10iya  & 0.224 & PTF & Cenko et al. 2012a \\           
SDSS\,J1342+0530$^1$ & 0.034 & SDSS & Wang et al. 2012 \\  
SDSS\,J1350+2916$^1$ & 0.078 & SDSS &  \\
PS1-10jh   & 0.170 & Pan-STARRS & Gezari et al. 2012 \\     
ASASSN-14ae & 0.044 & ASAS-SN & Holoien et al. 2014 \\
PTF\,09ge   & 0.064 & PTF & Arcavi et al. 2014 \\
PTF\,09axc$^{2}$  & 0.115 & PTF &  \\
PTF\,09djl  & 0.184 & PTF &  \\
PTF\,10nuj$^{3}$ & 0.132 & PTF & \\
PTF\,11glr$^{3}$ & 0.207 & PTF & \\ 
PS1-11af & 0.405 & Pan-STARRS & Chornock et al. 2014 \\
\hline
\end{tabular}
\caption{Candidate TDEs identified from X-ray, UV and optical observations. [Only sources
are listed which showed high-amplitude variability (largest in X-rays, and at least a factor
of 2 in the optical and UV), along with a good optical galaxy counterpart
for which follow-up spectroscopy confirmed an essentially in-active host galaxy. Most of the listed sources,
especially those with long-term coverage, are also quiescent in terms of their low-state X-ray emission
and absence of radio emission.   Also note that several events were detected in more than one waveband;
the sorting in the table is according to the waveband in which the event was initially identified.]  
\newline
\noindent $^1$ Identified by their transient emission lines. 
$^2$ Luminous X-ray emission 
of $7\,10^{42}$ erg/s was detected with \textit{Swift} in 2014 (Arcavi et al. 2014) well within
the regime of Seyfert galaxies, and therefore either indicates
long-lived AGN activity, or is still related to the optical flare.
$^3$ The positions of these two events are consistent with an off-center or a central location.    
}
\label{tab1-tdelist}
\end{center}
\end{table*}

Luminous, transient high-ionization lines with ionization potential
in the soft X-ray regime implied the presence of a high-amplitude high-energy outburst
(even though simultaneous X-ray observations did not exist).  
The most luminous coronal lines are those from
SDSSJ095209.56+214313.3. The highest ionization lines have faded by a factor $>10$ 
several years after high-state and are no longer detectable. The broad H$\alpha$ line 
is asymmetric and comes with an initial redshift of 600 km/s, and additional multi-peaked
narrow components are
present (Komossa et al. 2008).  
The highest-ionization lines, as well as broad H and He lines,
continue fading over several years, while the lower-ionization
lines like [OIII] increase in strength (Komossa et al. 2009, Wang et al. 2012,
Yang et al. 2013), consistent with light travel time effects, i.e., flare emission
reaching more extended regions of gaseous material/ISM  as time goes by.   

The broad emission lines from H and He may arise from the tidal debris itself
which forms the accretion disk
(see Guillochon et al. 2014, Bogdanovic et al. 2014,  
and Gaskell \& Rojas Lobos 2014 for models of PS1-10jh), while line emission from the
unbound stellar streams is expected to be very faint. 
Repeated spectral coverage shortly after peak is available for two of the events 
identified in the PTF archive (Arcavi et al. 2014), and one event identified in
the ASAS-SN survey (Holoien et al. 2014) along with optical lightcurves.
The Balmer emission lines show complex, variable profiles 
with kinematic shifts as high as several 1000 km/s. 

Likely, all events are related, and those with fainter broad lines
present a later stage in evolution (while strong narrow high-ionization
lines only appear
in gas-rich environments). A future test of the TDE
interpretation will consist of searching for similar events at off-nuclear locations.
While recoiling SMBHs and ongoing galaxy mergers 
will produce off-nuclear TDEs (Komossa \& Merritt 2008, Liu \& Chen 2013), 
these events would be rare in
the local universe, and the vast majority of nearby TDEs should therefore arise 
from the very cores of their host galaxies.

\section{\textit{Swift} discovery of jetted TDEs} 

\subsection{Observations of Swift\,J164449.3+573451} 

A new chapter in the study of TDEs was opened when \textit{Swift} detected the unique transient
GRB110328A/ Swift\,J164449.3+573451 (Swift\,J1644+57 hereafter).
It was first noticed when it triggered the \textit{Swift} Burst Alert Telescope (BAT) in March 2011
(Cummings et al. 2011), and initially
shared similarities with a GRB. However, the X-rays by far
did not fade as quickly as expected for GRBs. 

The rapid rise, huge X-ray peak luminosity, long duration, compact and variable associated radio emission,
and optically inactive host galaxy all contributed to the interpretation 
of this event as the launch of a powerful
jet following the tidal disruption of a star
(Bloom et al. 2011, Burrows et al. 2011, 
Zauderer et al. 2011, Levan et al. 2011).

The early evolution of the lightcurve is characterized by recurrent high-amplitude flares with
(isotropic) peak luminosities exceeding 10$^{48}$ erg/s (Burrows et al. 2011). After the first few days,
the lightcurve shows an overall decline (Fig. 6, 7), but continues to be complex and highly variable, with
variability timescales as short as $\sim$100 s.

Optical imaging and spectroscopy revealed the host galaxy at redsift $z=0.353$ (Levan et al. 2011).
The optical emission lines imply that the host is not an AGN, but an HII-type galaxy. Variable
emission was also detected in the NIR (not in the optical; likely due to the excess extinction
seen in the optical spectrum).
Various estimates of the SMBH mass, based on scaling relations or the most rapid variability timescale
all give $M_{\rm BH}$ $<$ $10^{7}$ M$_{\rm \odot}$.

\begin{figure}[t]
 \includegraphics[width=8.9cm]{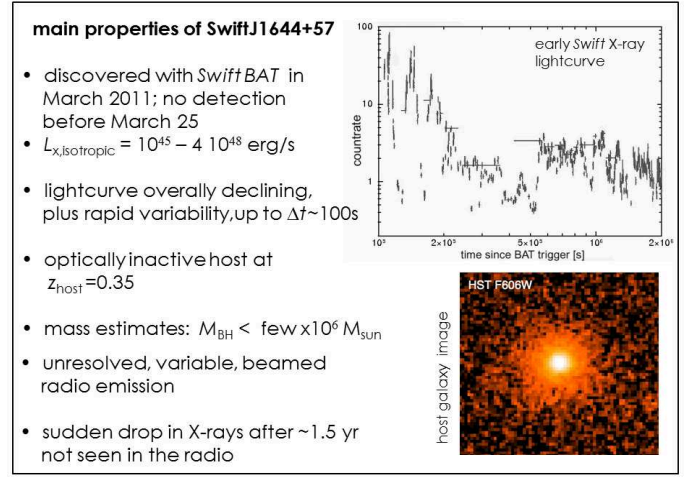}
\caption{Summary of some main observational properties of Swift\,J1644+75.
The panels show the early-phase X-ray
lightcurve (see Burrows et al. 2011) and the \textit{HST} optical image of the host
galaxy from Levan et al. (2011). 
}
\end{figure}

Swift\,J1644+57 was also associated with bright, compact, variable radio synchrotron emission 
(e.g., Zauderer et al. 2011) coincident with the nucleus of the host galaxy. Radio emission
was already 
detected during the first radio follow-up observation of Swift\,J1644+57
four days after the 
first X-ray detection with \textit{Swift}.  

Tidal disruption of a star powering a relativistic jet pointing in our
direction has been the widely favored explanation of this unique event.
Given the relatively low mass of the SMBH, the observed peak luminosity of Swift\,J1644+57 is up to a few
orders of magnitude above the Eddington limit, implying that either the accretion
at peak was highly super-Eddington, and/or that the emission was boosted due to a collimated relativistic jet.

A variety of follow-up observations of Swift\,J1644+57 have been carried out at all wavelengths 
(X-rays: Saxton et al. 2012, Reis et al. 2012, Castro-Tirado et al. 2013, Zauderer et al. 2013,
Gonzales-Rodriguez
et al. 2014, Mangano et al. 2015; radio: Berger et al. 2012, Zauderer et
al. 2013, Cendes et al. 2014; IR and radio polarimetry: Wiersema et al. 2012). Non-detections
were reported with \textit{VERITAS} and \textit{MAGIC} at energies $>$100 GeV (Aliu et al. 2011, 
Aleksic et al. 2013). 

In X-rays, evidence for a quasi-periodicity of 200\,s was reported by
Reis et al. (2012; see also Saxton et al. 2012). 
Unlike the long-term X-ray lightcurve, the radio emission from Swift\,J1644+75 continued to rise 
(Berger et al. 2012, Zauderer
et al. 2013; our Fig. 8).  
The \textit{Swift} X-ray lightcurve of Swift\,J1644+57 continued declining (Fig. 7), 
but then showed a sudden drop at $\sim$500 days
by a factor of $\sim$170; no longer detectable with
\textit{Swift} but still with \textit{Chandra} (Zauderer et al. 2013) and 
\textit{XMM-Newton} in deeper pointings  
(Gonzalez-Rodriguez et al. 2014).
The rapid decline is not seen in the radio (Fig. 8), 
implying that X-ray and radio emission
have different sites of origin
also at late times.
Zauderer et al. (2013) interpreted the drop in X-ray emission
as evidence for a change in accretion mode, turning off the jet production; while the faint
late-stage X-rays themselves (and the ongoing radio emission)
are consistent with arising from the forward shock related to the jet.

\begin{figure}[t]
\includegraphics[width=8.8cm]{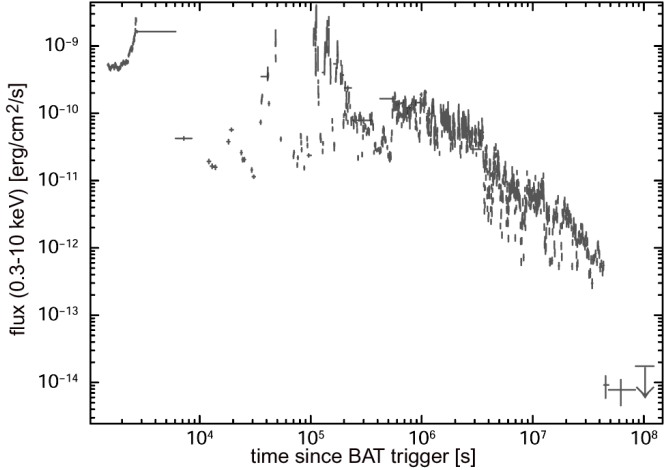}
\caption{\textit{Swift} long-term lightcurve of Swift\,J1644+75. 
Data are available at \textit{www.swift.ac.uk/xrt$_{-}$curves}. (See also Zauderer et al. 2013,
and see Mangano et al. 2015 for a detailed spectral and timing analysis of the whole \textit{Swift} 
lightcurve.)} 
   \label{fig3}
\end{figure}

\begin{figure}
\includegraphics[width=7.3cm]{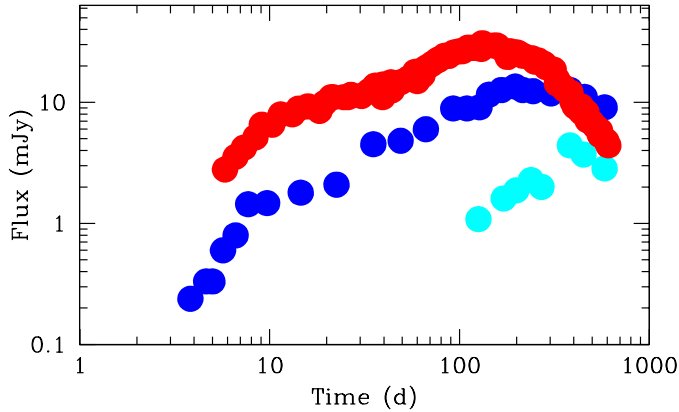}
\caption{Radio lightcurve (Zauderer et al. 2013) of Swift\,J1644+75, provided by A. Zauderer.
Frequencies of 1.8 GHz (cyan), 4.9 GHz (blue) and 15 GHz (red) are shown. 
}
\end{figure}

\subsection{Theory and implications} 

The discovery of Swift\,J1644+57 has triggered a large number of theoretical studies, addressing
the trajectory and type of the disrupted star and mass and spin of the SMBH 
(Shao et al. 2011, Miller \& G{\"u}ltekin 2011, Lei \& Zhang 2011, Cannizzo et al. 2011,
Krolik \& Piran 2011, Abramowicz \& Liu 2012), 
the nature of the radio
emission and general implications for the formation of radio 
jets (Metzger et al. 2012, Wang \& Cheng 2012, Krolik \& Piran 2012, De Colle et al. 2012, 
Cao \& Wang 2012, Gao 2012, Barniol Duran \& Piran 2013, Kumar et al. 2013, Zou et al. 2013,
Tchekhovskoy et al. 2014, Wang et al. 2014, Kelley et al. 2014, Parfrey et al. 2015,   
Liu et al. 2015, Mimica et al. 2015; see also comments by van Velzen et al. 2011b,
Giannios \& Metzer 2011), 
the evolution of the accretion
disk including effects of precession (Stone \& Loeb 2012b, Socrates 2012, Lei et al. 2013, 
Kawashima et al. 2013, Coughlin \& Begelman et al. 2014, Shen \& Matzner 2014), 
and a few alternative interpretations of the 
event (Ouyed et al. 2011, Quataert \& Kasen 201, Woosley \& Heger 2012).

A key question which has been raised (first mentioned in application 
to Swift\,J1644+57 by Bloom et al. 2011),
regards the role of magnetic fields in magnetohydrodynamic models of jet formation, and specifically
the question, whether the required large-scale magnetic field  
is generated \textit{in situ} in the disk, or is rather advected in with the flow. 
The magnetic field strength estimated for Swift\,J1644+57 is much higher than
that of a main-sequence star, and must have then been generated locally in the disk,
or requires the presence of a fossil disk.    

Regarding the source of the long-term radio emission there is overall concencus that it
is very likely synchrotron radiation from the shock which is formed when the jet interacts
with the interstellar medium.   
 
Another focus of attention has been the nature of the X-ray emission and its rapid variability
and early epochs of high-amplitude flaring. Most models have linked it with dissipation
in the inner jet (and/or effects of jet precession or nutation), 
and/or the forward shock, rather than directly with emission
from the accretion disk.  

Finally, a topic of close consideration has been the trajectory and type
of the disrupted star. While several studies have assumed, or explored, the disruption
of a main-sequence star, arguments have also been made in favor of the disruption
of a white dwarf (Krolik \& Piran 2011), a giant star (Shao et al. 2011),
or a star with a deeply plunging orbit (Cannizzo et al. 2011).   

\subsection{Swift\,J2058.4+0516} 

Evidence for a possible second jetted TDE discovered by \textit{Swift} was reported by Cenko et al (2012b). 
Swift\,J2058.4+0516 (Swift\,J2058+05 hereafter) exhibited a luminous, long-lived X-ray outburst
with an (isotropic) peak luminosity of $L_{\rm x} \approx 3\,10^{47}$ erg/s. Its X-ray lightcurve
shows rapid variability as fast as 1000s. The event is accompanied by strong radio emission. 
The likely host galaxy of Swift\,J2058+05 is at redshift $z=1.185$, and is optically inactive.
Because of the many similarities with Swift\,J1644+57, Cenko et al. suggested a similar outburst
mechanism, consistent with multi-wavelength follow-up observations (Pasham et al. 2015). Notably,
like Swift\,J1644+57, the X-ray lightcurve of Swift\,J2058.4+0516 shows an abrupt drop (between
days 200-300). Since the X-rays vary rapidly before the drop, an origin near the SMBH rather than
in a forward shock is implied (Pasham et al. 2015).      

\subsection{Radio emission from other TDEs} 

The possibility that TDEs launch radio jets, came up
with the detection of the first few X-ray TDEs with \textit{ROSAT}. 
In order to test, whether TDEs do launch radio jets (and to exclude
a peculiar mini-blazar at the galaxy core), radio follow-ups
of NGC\,5905 were performed with the VLA
(Komossa 2002) 6 years after the initial X-ray outburst.
No radio emission was detected ($f_{\rm R} < 0.15$ mJy at 8.46 GHz; see
Sect. 2.3 of Komossa 2002 for an extended discussion of the radio properties
of NGC\,5905).

More recently, deep radio follow-ups have been carried out of some
of the newly identified (candidate) TDEs (e.g., van Velzen et al. 2011,
Saxton et al. 2012, Cenko et al. 2012, Saxton et al. 2014,  
Chornock et al. 2014). No radio emission
was detected. Radio observations were also carried out for several of the previously
known TDEs, covering time intervals of up to more than a  decade (the \textit{ROSAT} events),
and years (Bower et al. 2013, van Velzen et al. 2013).  Most sources were undected with
the VLA in either study, with upper limits on the flux density in the range $f_{\rm R} < 10-200 \mu$Jy.
The two exceptions (Bower et al. 2013) are the Seyfert galaxy IC\,3599 
[the radio emission in this AGN is most likely linked to the 
long-lived Seyfert activity of this galaxy,
and the nature of its (repeat) X-ray flaring remains debated; 
e.g., Komossa et al. 2015, see our Sect. 7] and a source
in the X-ray error circle of the \textit{ROSAT} event RXJ1420+5334. 

Jets similar to the one of Swift\,J1644+57 should have been detected in
the deepest radio follow-ups, even if not pointed at us. 
Therefore, most TDEs do not launch powerful radio jets. 

The fact that Swift\,J1644+57 does, has re-raised the question, which
parameter(s) determine jet formation. In application to  Swift\,J1644+57, 
BH spin or the pre-existence of a fossil disk or other mechanisms
to produce a large magnetic flux to power the jet have been considered
(Sect. 5.2).

\section{TDE rates}

Several approaches have been followed to estimate TDE rates from observations, making use of the
\textit{ROSAT} all-sky survey (Donley et al. 2002, 	
Khabibullin \& Sazonov 2014), 
the \textit{XMM-Newton} slew survey (Esquej et al. 2008),
upper limits from \textit{Chandra} deep fields (Luo et al. 2008),
clusters of galaxies (Maksym et al. 2010),
GALEX (Gezari et al. 2008), and SDSS (e.g., Wang et al. 2012, van Velzen \& Farrar 2014). 
All rates are in the range $10^{-4}-10^{-5}$/yr/galaxy,
and consistent with order-of-magnitude theoretical predictions which
are on the same order 
(e.g., Rees 1990, Merritt 2009, review by Alexander 2012).  

\section{TDEs in AGN?} 

TDEs might well occur in AGN, too, and rates are expected to be high (Karas \& Subr 2007). 
However, it is generally more challenging to identify TDEs in AGN because of the much lower contrast
of any TDE-related accretion flare relative to the permanently bright accretion disk. Further,
it is more difficult to uniquely associate any particular flare in classical AGN (i.e., galaxies 
which show long-lived activity in X-rays, and/or radio, and characteristic bright 
emission-lines from
the narrow-line region) with a TDE, because processes in the long-lived 
accretion disk itself can potentially cause
high-amplitude variability.{\footnote{At the same time, a TDE in a gas rich environment
like a starburst galaxy will inevitably excite emission lines which temporarily look
like an AGN (e.g., Komossa et al. 2008). Therefore, repeat optical spectroscopy 
after a luminous flare is required, in order
to assess whether any narrow lines are persistent or vary and ultimately fade.}}  
Nevertheless, a few cases of unusually high-amplitude variability of 
AGN have been considered in the context of 
a tidal disruption scenario. 

\subsection{The Seyfert galaxy IC\,3599} 
(1) The Seyfert galaxy IC\,3599
showed an  X-ray outburst during the \textit{ROSAT} all-sky
survey, accompanied by a temporary remarkable brightening and fading 
of the optical emission lines
(Brandt et al. 1995, Grupe et al. 1995a,
Komossa \& Bade 1999). The cause of the outburst remained unknown,
and   
high-amplitude narrow-line Seyfert 1 (NLS1) variability{\footnote{we note in passing,
that IC\,3599 is not a classical NLS1 galaxy, though, since it lacks
the strong, characteristic FeII emission, typical for these galaxies,
and since the only available optical high-state spectrum (Brandt et al. 1995) was
just a snapshot of the fading broad emission lines responding to the flare, and the
full line width can therefore be much higher than the value of 2000 km/s typically
used to define NLS1 galaxies as a class}}, a disk instability, or a 
TDE have all been considered. IC\,3599 was already known
as an active galaxy before the outburst (based on its narrow emission lines like [OIII]). 
Recently, a second outburst of IC\,3599 was discovered by \textit{Swift} in 2010 
(Komossa et al. 2015,
Campana et al. 2015,
Grupe et al. 2015a). Recurrent flaring within decades is uncommon in TDE
scenarios, even though not impossible, since, for instance, rates are strongly 
boosted under some conditions in galaxy mergers  
(e.g., Chen et al. 2009), and repeat tidal stripping 
can occur for orbiting stars surviving the first encounter.
This latter possibility was explored by
Campana et al. (2015), while processes related to the AGN accretion disk 
(and SMBBHs) where
discussed by Komossa et al. (2015) and Grupe et al. (2015a).  
Variability of similar magnitude as IC\,3599 (a factor $\sim100$) is rare, but has also been
observed in a few other AGN{\footnote{e.g., 1E1615+061 (factor 100, Piro et al. 1988),
WPVS\,007 (factor 400, Grupe et al. 1995b, Vaughan et al. 2004, Grupe et al. 2013),
PHL\,1092 (factor 200, Miniutti et al. 2009), GSN\,069 (factor 240, Miniutti et al. 2013), 
and RXJ2317$-$4422 (factor 60, Grupe et al. 2015b)}}, likely related
to processes in the long-lived AGN accretion disk (and this 
therefore is a likely explanation of IC\,3599, too), or to
absorption.

\subsection{Other cases in the UV and X-rays} 

(2) In the UV, Renzini et al. (1995) observed faint, enhanced emission from the core of 
the elliptical radio galaxy NGC\,4552 
with \textit{HST}, and suggested a mild accretion event from an interstellar 
cloud or the tidal stripping of a star; 
or fluctuations
in the accretion rate of the very faint AGN in NGC\,4552 (Cappellari et al. 1999, Renzini 2001). 
The UV spike varied by a factor of $\sim4.5$
on the timescale of several years.  Very generally, Maoz et al. (2005) found, that 
nuclear UV variability in LINER galaxies with radio cores is very common.  
(3) Meusinger et al. (2010) observed a high-amplitude (factor 20) UV flare from the quasar J004457+4123,
which they interpreted as either a microlensing event or else a TDE.  
(4) Nikolajuk \& Walter (2013) reported the discovery of a hard X-ray flare with \textit{INTEGRAL},
associated with the Seyfert galaxy NGC\,4845.  
Because of the high amplitude of variability, the authors 
strongly favored a tidal disruption scenario; in this
case the 
disruption of a giant planet.
(5) The galaxy XMM\,J061927.1$-$655311 was found in a high-amplitude flaring 
state (factor $>$140) with \textit{XMM-Newton} (Saxton et al. 2014) at 
$L_{\rm x,peak} \sim 10^{44}$ erg/s.
Optical follow-up spectroscopy revealed low-level
AGN activity (even though the spectrum looks peculiar and narrow emission lines
are very faint). The X-rays and UV, observed with \textit{Swift}, declined  
subsequently. The event was interpreted as either a TDE or a change in accretion rate
in a long-lived AGN.  

\section{Future TDE surveys}

Upcoming sky surveys will find TDEs in the hundreds or thousands, including in the radio
with SKA (Donnarumma et al. 2015), in the optical with LSST 
(van Velzen et al. 2011),
and in hard X-rays with LOFT (Rossi et al. 2015). Further, a dedicated X-ray
transient mission is under consideration in China, which has among its prime
goals the detection and study of TDEs in large numbers; the soft X-ray mission
\textit{Einstein Probe} (Zhao et al. 2014, Yuan et al. 2015). 
\textit{Einstein Probe} is designed to carry out an all-sky transient survey
at energies of 0.5--4 keV. The concept is based on a wide-field micro-pore Lobster-eye
imager (60$^{\rm o}$ $\times$ 60$^{\rm o}$), and a more sensitive narrow-field instrument for follow-ups.
The large number of new events and the well-covered
lightcurves will enable a wealth of new science. 
In particular, X-rays will be sensitive to relativistic
effects and accretion physics near the last stable orbit.

Meanwhile, \textit{Swift} will continue to greatly contribute to this field, including searching
for the re-emergence of bright X-rays from Swift\,J1644+57 itself, predicted by some models.  

\vskip0.3cm

\noindent {\bf{Acknowledgments}}

{It is my pleasure to thank Andrew Levan, Shuo Li, and Ashley Zauderer 
for providing figures,
Fukun Liu for numerous discussions on TDEs, Eduardo Ros 
for a critical reading of the manuscript, and the anonymous referee 
for useful comments. 
I would like to thank the
Aspen Center for Physics for support and hospitality during a workshop
in 2012, and the participants for illuminating discussions on TDEs. 
The Aspen Center for Physics is
supported by the National Science Foundation under grant No. PHYS-1066293. }





\end{document}